\documentclass[12pt]{article}

\usepackage{amsfonts}

\RequirePackage{times}
\RequirePackage{mathptm}

\newtheorem{definition}{{\bf Definition} }

\begin{document}

\title{How to acknowledge hypercomputation?}
\author{Alexander Leitsch\footnote{leitsch@logic.at}, G{\"{u}}nter Schachner \\
 {\small Institut f{\"{u}}r Computersprachen, Vienna University of Technology,}\\ {\small   Favoritenstr.9/185, 1040 Vienna, Austria}\\
{and}\\
Karl Svozil\footnote{svozil@tuwien.ac.at} \\
 {\small Institute for Theoretical Physics, Vienna University of Technology,}\\ {\small   Wiedner Hauptstra\ss e 8-10/136, 1040 Vienna, Austria}
      }
\date{ }
\maketitle


\begin{abstract}
We discuss the question of how to operationally validate whether or not a ``hypercomputer'' performs better than the known discrete computational models.
\end{abstract}

\section{Introduction}

It is widely acknowledged \cite{wolfram-2002,svozil-2005-cu} that every physical system corresponds to a computational process,
and that every computational process, if applicable, has to be physically and operationally feasible in some concrete realization.
In this sense, the physical and computational capacities should match;
because if one is lagging behind the other,
there is a lack in the formalism and its potential scientific and ultimately technological applicability.
Therefore, the exact correspondence of the mathematical formalism on the one hand, and
the particular physical system which is represented by that formalism  on the other hand, demands careful attention.

If one insists on operationalizability
\cite{bridgman},
one needs not go very far in the history of mathematics to encounter
suspicious mathematical objects.
Surely enough, the number $\pi$ can be defined and effectively computed as the ratio of the
circumference to the diameter of a ``perfect (platonic)'' circle.
Likewise, the numbers $\sqrt{2}$ and $\sqrt{3}$ can be interpreted as the ratio
between the length of the diagonal to
the side length of any square and cube.
But it is not totally unjustified to ask whether or not these numbers have
any operational meaning
in a strict physical sense; i.e., whether such numbers could, at least in
principle, be constructed and measured with arbitrary or even with absolute
precision.

At the heart of most of the problems seems to lie the ancient issue of the ``very large/small'' or even potential
infinite versus the actual infinite.
Whereas the mathematical formalism postulates the existence of  actual
infinite constructions and methods,
such as the summation of a (convergent) geometric series, or diagonalization,
the physical processes, methods and techniques are never infinite.

Suppose, as an example, one would attempt to operationalize $\pi$.
Any construction of a ``real'' circle and one of its diameters, and a
subsequent measurement thereof,
would find its natural scale bound from below by the atomistic structure of
matter upon which any such  circle
is based.
Long before those molecular or atomic scales,
the physical geometry might turn out to be not as straightforward as it
appears from
larger scales; e.g., the object might turn out to be a fractal.

Chaitin's Omega number \cite{chaitin3} is
interpretable as the halting probability of a universal computer.
Omega is provably uncomputable; i.e., no computer, in any of the (equivalent universal) forms conceptualized so far,
``computes'' Omega. Thus, only hypercomputers might be capable to ``compute'' Omega.
Yet, Omega can be ``computed in the limit'' (without any computable radius of convergence)
by a finite-size program in infinite time and with infinite space.
Just as for many transcendential numbers --- the difference being the absence of any computable radius of convergence ---
the first digits of Omega are well known
\cite{calude-dinneen06}, yet
Omega is provable algorithmically incompressible and thus random.
Nevertheless, presently, for all practical purposes, the statement that
``the $10^{10^{10^{10}}}$th digit in a decimal expansion of Euler's number $e$ is $5$''
is as unverifiable as a similar statement for Omega.
[Note, however, that for certain transcendental numbers  such as Pi,
the Bailey--Borwein--Plouffe (BBP) formula \cite{bailey97,bailey05}
can be used to efficiently calculate ``large'' digits
in various bases without knowing all of the previous digits.]
Omega encodes all decision problems which can be algorithmically interpreted.
For instance, for a particular universal computer,
Goldbach's conjecture and Riemann's hypothesis could be decided with programs of size 3484 and 7780 bits,
respectively \cite{calude-elena-dinneen06}.
Yet, Omega  appears to have two features which are normally considered contradictory:
it is one of the most informative mathematical numbers imaginable.
Yet at the same time this information is so compressed that it cannot be deciphered;
thus Omega appears to be totally structureless and random.
In this sense, for Omega, total information and total randomness seem to be ``two sides of the same coin.''
On a more pragmatic level, it seems impossible here to differentiate between order and chaos,
between knowledge and chance.
This gives a taste of what can be expected from any ``hypercomputation''
beyond universal computability as defined by Turing.

It should always be kept in mind that all our sense perceptions are derived from elementary discrete events,
such as clicks in photon or particle detectors, even if they appear to be
analog:
the apparently smooth behavior has a discrete  fine structure.
Among other issues, such as finiteness of system resources,
this discreteness seems to prohibit the ``physical realization'' of any actual infinities.

What is the physical meaning of infinite  concepts,
such as space-time singularities, point particles, or infinite precision?
For instance, are infinity machines with geometrically squeezed time cycles,
such as the ones envisioned by Weyl \cite{weyl:49}
and others \cite{gruenbaum:74,thom:54,benna:62,rucker,pit:90,ear-nor:93,hogarth1,hogarth2,beth-59,le-91,sv-aut-rev}
physically feasible?
Motivated by recent proposals to utilize quantum computation for trespassing the Turing barrier
\cite{2002-cal-pav,ad-ca-pa,kieu-02,kieu-02a},
these accelerating Turing machines have been intensively discussed \cite{ord-2006}
among other forms of hypercomputation \cite{Davis-2004,Doria-2006,Davis-2006}.

Certainly, the almost ruthless and consequential application of seemingly
mind-boggling theories
such as quantum mechanics, as  far as finitistic methods are concerned, has
yielded one success after another.
Nevertheless, the use of actual
transfinite concepts and methods remains highly conjectural.

A priori, while it may appear rash to exclude the transfinite in general, and transfinite set theory
in particular from physics proper,
one should be aware of its counterintuitive consequences, such as for instance
the Banach-Tarski paradox, and be careful in claiming its physical
applicability.
Recall the old phrase attributed to Einstein and Infeld
(Ref.~\cite{ein-in}, p.31),
{\em ``Physical concepts are free creations of the human
   mind, and are not, however it may seem,
   uniquely determined by the external world.''}

To this point, we are not aware of any test, let alone any application, of the actual
transfinite in Nature.
While general contemplations about hypercomputations and the applicability of transfinite concepts
for physics may appear philosophically interesting,
our main concern will be operational {\em testability}:
if presented with claims that hypercomputers exist, how could we possibly falsify, or even verify and test such propositions
\cite{Chow-2004}?

In what follows, hypercomputation will be conceptualized in terms of a black box with its input/output behavior.
Several tests and their rather limited scope will be evaluated.
Already in 1958, Davis \cite[p. 11]{davis-58}
sets the stage of the following discussion by pointing out
 {\em `` $\ldots$ how can we ever exclude the possibility of our being presented,
 some day (perhaps by some extraterrestrial visitors), with a (perhaps
 extremely complex) device or ``oracle'' that ``computes'' a
 non-computable function?''}
While this may have been a remote, amusing issue in the days written,
the advancement of physical theory in the past decades
has made necessary
a careful evaluation of the possibilities and options for
verification and falsification of certain claims that a concrete physical system
``computes''   a  non-computable function.

\section{On black boxes which are hypercomputers}

Digital computers  are ``reliable'' in the sense of ``comprehensible, reproducible, traceable, accountable,''
just for the reason that the internal functioning of a
digital computer is known to anyone seeking such knowledge, and that the programs give a logical
specification of an algorithm. Because the code is known one can -- at least in
principle -- verify the behavior for {\em all} inputs mathematically.

\begin{figure}
\begin{center}
\unitlength 0.6mm
\linethickness{0.4pt}
\begin{picture}(79.33,115.00)
\put(2.33,14.67){\framebox(5.00,5.00)[cc]{{\tiny $ i_{1}$}}}
\put(9.33,14.67){\framebox(5.00,5.00)[cc]{{\tiny  $i_{2}$}}}
\put(16.33,14.67){\framebox(5.00,5.00)[cc]{{\tiny $i_{3}$}}}
\put(23.33,14.67){\framebox(5.00,5.00)[cc]{{\tiny $i_{4}$}}}
\put(30.33,14.67){\framebox(5.00,5.00)[cc]{{\tiny $i_{5}$}}}
\put(37.33,14.67){\framebox(5.00,5.00)[cc]{{\tiny $i_{6}$}}}
\put(44.33,14.67){\framebox(5.00,5.00)[cc]{{\tiny $i_{7}$}}}
\put(51.33,14.67){\framebox(5.00,5.00)[cc]{{\tiny $i_{8}$}}}
\put(58.33,14.67){\framebox(5.00,5.00)[cc]{{\tiny $i_{9}$}}}
\put(65.33,14.67){\framebox(5.00,5.00)[cc]{{\tiny $i_{10}$}}}
\put(72.33,14.67){\framebox(5.00,5.00)[cc]{{\tiny $i_{11}$}}}
\put(2.33,7.67){\framebox(5.00,5.00)[cc]{{\tiny $i_{12}$}}}
\put(9.33,7.67){\framebox(5.00,5.00)[cc]{{\tiny $i_{13}$}}}
\put(16.33,7.67){\framebox(5.00,5.00)[cc]{{\tiny $i_{14}$}}}
\put(23.33,7.67){\framebox(5.00,5.00)[cc]{{\tiny $i_{15}$}}}
\put(30.33,7.67){\framebox(5.00,5.00)[cc]{{\tiny $i_{16}$}}}
\put(37.33,7.67){\framebox(5.00,5.00)[cc]{{\tiny $i_{17}$}}}
\put(44.33,7.67){\framebox(5.00,5.00)[cc]{{\tiny $i_{18}$}}}
\put(51.33,7.67){\framebox(5.00,5.00)[cc]{{\tiny $i_{19}$}}}
\put(58.33,7.67){\framebox(5.00,5.00)[cc]{$\cdot$}}
\put(65.33,7.67){\framebox(5.00,5.00)[cc]{$\cdot$}}
\put(72.33,7.67){\framebox(5.00,5.00)[cc]{$\cdot$}}
\put(20.00,35.00){\line(1,0){30.00}}
\put(50.00,35.00){\line(0,1){25.00}}
\put(50.00,60.00){\line(-1,0){30.00}}
\put(20.00,60.00){\line(0,-1){25.00}}
\put(50.00,60.00){\line(3,2){15.00}}
\put(20.00,60.00){\line(3,2){15.00}}
\put(50.00,35.00){\line(3,2){15.00}}
\put(65.00,45.00){\line(0,1){25.00}}
\put(65.00,70.00){\line(-1,0){30.00}}
\put(0.00,5.00){\line(1,0){79.33}}
\put(79.33,5.00){\line(0,1){17.00}}
\put(79.33,22.00){\line(-1,0){79.33}}
\put(0.00,22.00){\line(0,-1){17.00}}
\put(30.00,90.00){\line(1,0){20.00}}
\put(50.00,90.00){\line(0,1){20.00}}
\put(50.00,110.00){\line(-1,0){20.00}}
\put(30.00,110.00){\line(0,-1){20.00}}
\multiput(40.67,66.33)(0.11,0.16){15}{\line(0,1){0.16}}
\multiput(42.35,68.80)(0.12,0.20){11}{\line(0,1){0.20}}
\multiput(43.67,70.99)(0.12,0.24){8}{\line(0,1){0.24}}
\multiput(44.62,72.90)(0.12,0.33){5}{\line(0,1){0.33}}
\multiput(45.20,74.55)(0.11,0.69){2}{\line(0,1){0.69}}
\multiput(45.42,75.92)(-0.08,0.55){2}{\line(0,1){0.55}}
\multiput(45.27,77.01)(-0.10,0.16){5}{\line(0,1){0.16}}
\multiput(44.75,77.84)(-0.18,0.11){5}{\line(-1,0){0.18}}
\multiput(43.87,78.39)(-0.42,0.09){3}{\line(-1,0){0.42}}
\put(42.62,78.66){\line(-1,0){1.62}}
\put(41.00,78.67){\line(-1,0){1.85}}
\multiput(39.15,78.71)(-0.50,0.08){3}{\line(-1,0){0.50}}
\multiput(37.66,78.95)(-0.28,0.11){4}{\line(-1,0){0.28}}
\multiput(36.53,79.41)(-0.13,0.11){6}{\line(-1,0){0.13}}
\multiput(35.78,80.07)(-0.10,0.22){4}{\line(0,1){0.22}}
\put(35.39,80.93){\line(0,1){1.07}}
\multiput(35.36,82.00)(0.11,0.43){3}{\line(0,1){0.43}}
\multiput(35.70,83.28)(0.12,0.25){6}{\line(0,1){0.25}}
\multiput(36.41,84.76)(0.12,0.19){9}{\line(0,1){0.19}}
\multiput(37.48,86.44)(0.12,0.15){24}{\line(0,1){0.15}}
\multiput(39.67,34.67)(0.18,-0.11){9}{\line(1,0){0.18}}
\multiput(41.31,33.67)(0.11,-0.13){8}{\line(0,-1){0.13}}
\put(42.18,32.59){\line(0,-1){1.16}}
\multiput(42.29,31.43)(-0.11,-0.21){6}{\line(0,-1){0.21}}
\multiput(41.64,30.18)(-0.14,-0.11){19}{\line(-1,0){0.14}}
\multiput(39.00,28.00)(-0.22,-0.12){8}{\line(-1,0){0.22}}
\multiput(37.23,27.05)(-0.12,-0.13){8}{\line(0,-1){0.13}}
\multiput(36.27,26.01)(-0.07,-0.57){2}{\line(0,-1){0.57}}
\multiput(36.12,24.86)(0.11,-0.21){6}{\line(0,-1){0.21}}
\multiput(36.79,23.62)(0.13,-0.12){14}{\line(1,0){0.13}}
\put(35.00,47.33){\makebox(0,0)[cc]{\huge ?}}
\put(40.00,100.00){\makebox(0,0)[cc]{$o_j$}}
\put(40.00,0.00){\makebox(0,0)[cc]{input interface}}
\put(40.00,115.00){\makebox(0,0)[cc]{output interface}}
\put(70.00,55.00){\makebox(0,0)[lc]{black box}}
\end{picture}
\end{center}
\caption{Hypercomputers of unknown origin may be studied phenomenologically;
i.e., by considering them as a black box with input and output interfaces, and by studying their input-output behavior.
\label{2007-hc-f1}}
\end{figure}
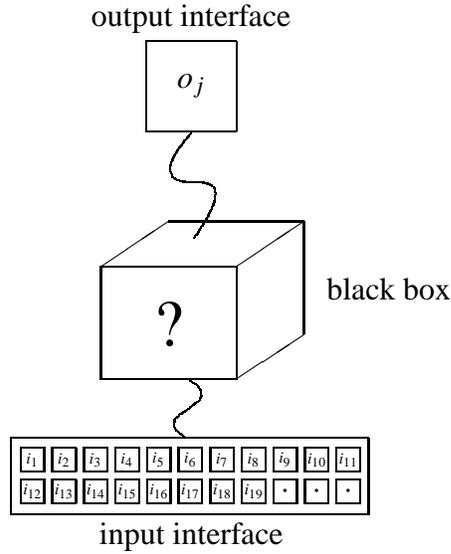

One cannot expect that hypercomputers will be ``reliable'' in the aforementioned sense.
The alleged  ``hypercomputer'' may be presented as a device, an agent or an oracle
of unknown provenance, and in particular with no
rational explanation (in the traditional algorithmic sense) of its intrinsic working.
(For the sake of an example, consider Laplace being confronted with a modern digital clock.)
Therefore, we attempt to deal with hypercomputers purely phenomenologically;
i.e., by analyzing their input/output behavior.
Thereby, the hypercomputer will be treated as black box with input and output interfaces; see Fig.~\ref{2007-hc-f1}.

The following  notation is introduced. Let $B$ be a subset of $X_1 \times \ldots
\times X_m$. The $i$-th projection of $B$ (for $i=1,\ldots , m$),
written as $B_i$, is defined by:
\[
\begin{array}{lcl}
B_i &=& \{x \mid x \in X_i, \ (\exists y \in X_1 \times \ldots
\times X_{i-1})(\exists z \in X_{i+1} \times \cdots \times X_m)\\
& & (y,x,z) \in B\}.
\end{array}
\]
For any $x \in N^m$ we define
$$|x| = \max\{x_i \mid i \in \{1,\ldots,m\}\}.$$

Then, a hypercomputer can be defined via its input/output behavior of black boxes as follows.

\begin{definition}[black box]\label{def.blackb}
Let $X,Y$ be sets and ${\Bbb N}$ be the set of natural numbers. A subset
$B$ of $X \times Y \times {\Bbb N}$ is called a {\em black box} if $B_1
= X$. $X$ is called the {\em input set} and $Y$ the {\em output
set.}
\end{definition}
Note that the condition $B_1 = X$ models a computing device
which is total, i.e. there exists always an output.
\begin{definition}
Let $B$ be a black box. We define
\begin{eqnarray*}
f_B &=& \{(x,y) \mid (\exists z)(x,y,z) \in B\},\\[1ex]
t_B &=& \{(x,z) \mid (\exists y)(x,y,z) \in B\}.
\end{eqnarray*}
$f_B$ is called the {\em input-output relation} of $B$ and $t_B$
the {\em computing time} of $B$. If $f_B$ and $t_B$ are functions
then $B$ is called deterministic.
\end{definition}
Every halting deterministic Turing machine defines a
black box. Indeed, let $M$ be a Turing machine (computing a total
function), $f_M\colon X \to Y$ be the function computed by $M$ and
$t_M$ be the computing time of $M$. Then
$$\{(x,f_M(x),t_M(x)) \mid x \in X\}$$
is a (deterministic) black box. Similarly all halting
non-deterministic Turing machines define black boxes.

\begin{definition}[hypercomputer]\label{def.hyperc}
A {\em strong hypercomputer} is a black box $B$ where $f_B$ is not
Turing-computable.
\end{definition}

\begin{definition}
Let ${\cal C}$ be a class of computable monotone functions ${\Bbb N} \to
{\Bbb N}$ containing the polynomials (over ${\Bbb N}$ with non-negative
coefficients). Then ${\cal C}$ is called a {\em bound class.}
\end{definition}

\begin{definition}
A {\em weak hypercomputer} is a black box $B$ with the
following property:
There exists a bound class ${\cal C}$ such that
\begin{itemize}
\item $t_M(x) > g(|x|)$ almost everywhere for all $g \in {\cal C}$ and for
    all Turing machines $M$ with $f_M =f_B$.
\item There exists an $h \in {\cal C}$ such that $t_B(x) \leq h(|x|)$ for
        all $x \in B_1$.
\end{itemize}
\end{definition}

A strong hypercomputer computes either a
non-computable function or decides an undecidable problem. A weak
hypercomputation outperforms all Turing machines. A possible
scenario for a weak hypercomputer $B$ is the following:
$f_B$ is an ${\bf EXPTIME}$-complete problem, therefore there exists no
polynomial $p$ and no Turing machine $M$ computing $f_B$ with
$t_M(x) \leq p(|x|)$ for all $x \in X$ , but $t_B(x) \leq p(|x|)$
for all $x \in X$ and for a polynomial $p$.

For non-deterministic hypercomputers we may distinguish between the following cases:
\begin{itemize}
\item $f_B$ is not a function,
\item $f_B$ is a function, but $t_B$ is not.
\end{itemize}

For stochastic hypercomputers, either $t_B$ or both $f_B$ and
$t_B$ are random variables, and the requirements on the computation have to be
specified.

\section{Tests}

Having set the stage for a general investigation into hypercomputers which are presented to us as
black boxes, we shall consider a few cases and tests.
These test methods will be essentially heuristic and present no way of systematically addressing
the issue of falsifying or even verifying hypercomputation.

One strategy for creating tests will be to consider problems which are {\em asymmetric} with respect to their
{\em creation} and {\em verification} --- which should be ``easy'' --- on the one hand,
and their {\em solution} --- which should be ``hard'' --- on the other hand.

\subsection{NP-complete cases}

It may be conjectured that, by operational means,  it is not possible to go beyond
tests of hyper-NP-completeness.
Even for an NP-complete problem as for instance the satisfiability problem of propositional logic (SAT), it is not trivial to
verify that a hypercomputer solves the problem in polynomial time.
Without insight into the internal structure of the hypercomputer
we cannot obtain a proof of polynomial time computation,
which is an asymptotic result. Even here we rely on
experiments to test a ``large'' number of
problems. A central problem consists in the right selection of
problem sequences. If the selection is based on random generators
we merely obtain results on average complexity, which would not be
significant.

Furthermore, we need at least some information about the polynomial
in question (e.g., its maximum degree). Otherwise it remains
impossible to decide by finite means whether some behavior
is polynomial or not.

\subsection{Harder cases with tractable verification}

Do there exist (decision) problems which are harder
than the known NP-complete cases,
possibly having no recursively enumerable solution and proof methods,
whose results nevertheless are tractable verifiable?
For example, the problem of \emph{graph non-isomorphism} (GNI) is one that is not
known to be in NP, not even in NP $\cup$ BPP. Nevertheless, it is
possible to ``efficiently verify'' whether a ``prover''
solves this problem correctly.

If the prover claims that two graphs $G_1$ and $G_2$ are isomorphic,
he can convince us by providing a graph isomorphism. That can be checked
in polynomial time, which also means that GNI $\in$ coNP.
If, on the other hand, the prover claims that $G_1$ and $G_2$ are
non-isomorphic, we can verify this by the following \emph{interactive proof}:

\begin{enumerate}
\item Choose one of the graphs $G_1$ and $G_2$ with equal probability. \label{I:1}
\item Apply an arbitrary permutation to its vertices; this yields graph $H$.
\item The prover must decide whether $H$ is equivalent to $G_1$ or $G_2$. \label{I:3}
\item Repeat for $N$ rounds.
\end{enumerate}

If the initial answer was wrong and the graphs $G_1$ and $G_2$ are actually isomorphic,
the prover can in step \ref{I:3} only \emph{guess} which graph
was chosen in step \ref{I:1} (since now $H$ could have been derived from
\emph{either}). Hence, after $N$ rounds we can be sure with probability
$1-2^{-N}$ that the graphs $G_1$ and $G_2$ are non-isomorphic.

By denoting the class of interactive proofs by IP, we have shown that
GNI $\in$ IP. Interactive proofs further exist for \emph{every} language in
PSPACE (which is assumed to be \emph{much} larger than NP). In fact, it can be shown
\cite{sha:92} that IP equals PSPACE. This means, in particular, that
IP is closed under complement.

The protocol in the example above has the property that in each round a constant number
of messages is sent. In a generic interactive proof system for
PSPACE this is \emph{not} necessarily true; but at any instance the number of messages depends
polynomially on the input length.

In the literature, specific classes of interactive proof systems are investigated
as well, e.g.\ the \emph{Arthur-Merlin class} \cite{bab:85} and
the \emph{Goldwasser-Micali-Rackoff (GMR)  class} \cite{GMR:85}. The former uses public coin tosses, with the
intention to accommodate certain languages in as low complexity classes as possible.
The latter uses private coin tosses, with the intention to cover the widest
possible class of efficiently verifiable languages;
additionally, it has the feature of providing \emph{zero-knowledge proofs,}
which is of great significance in cryptography. (The protocol presented above does
not have the zero-knowledge property -- unless GNI $\in$ BPP --, but can be modified
to have.) For further information on interactive proof systems see \cite{BM:88,gold:01}.

\subsection{Interference of problems}

One may confront the hypercomputer with the problem
of comparing the solutions of multiple tasks.
Such a comparison needs not necessarily involve the separate computation of the solutions of these multiple tasks.

As an analogy, consider Deutsch's problem as one of the first problems which
quantum computers could solve effectively.
Consider a function that takes a single (classical) bit into a single (classical) bit.
There are four such functions $f_1,\ldots ,f_4$, corresponding to all variations.
One can specify or ``prepare'' a function bitwise, or alternatively,
one may specify it by requiring that, for instance, such a function
acquires different values on different inputs, such as $f(0)\neq f(1)$.
Thereby, we may, even in principle, learn nothing about the individual functional values alone.

A related issue is the question of whether or not one hypercomputer should be allowed to verify another hypercomputation.
This would be similar to using one algorithm to prove the four color theorem,
and, since it is unfeasible for a human to verify by hand,
another algorithm to check the validity of the former computer-assisted proof.

\subsection{Generation of random sequences}
By implementation of Chaitin's ``algorithm'' to compute
Chaitin's $\Omega$  \cite{chaitin:01}
or variants thereof \cite{calude:94},
it would in principle be possible to ``compute'' the first bits of random sequences.
Such random sequences could in principle be subject to the usual
tests of stochasticity \cite{svozil-qct,calude-dinneen05}.

Note that in quantum mechanics, the randomness of certain microphysical events,
such as the spontaneous decay of excited quantum states
\cite{erber-95,berkeland:052103},
or the quantum coin toss experiments in complete context mismatches
\cite{svozil-qct} is postulated as an axiom.
This postulate is then used as the basis for the production of quantum randomness oracles
such as the commercially available {\it Quantis}\textsuperscript{\texttrademark} interface \cite{Quantis}.

\section{Impossibility of unsolvable problems whose ``solution'' is polynomially verifiable}
Let $\Sigma_0$ be a finite (non-empty) alphabet and $X \subset \Sigma_0^*$
be a semi-decidable, but not decidable set. That means there exists a
Turing machine which accepts
the language $X$, but does not terminate on all $x \in \Sigma_0^*$.
The concept
of {\em acceptable by Turing machines} is equivalent to {\em derivable by
inference systems} or {\em producible by grammars}. We choose the approach of
a {\em universal proof system}, i.e. of a system which simulates every
Turing machine.

Let $P$ be such a proof system. Let $V$ be an infinite set of variables
(over strings in $\Sigma^*$). A {\em meta-string} is an object
$x_1 \ldots x_n$ where
$x_i \in \Sigma$ or $x_i \in V$. If $X$ is a meta-string and $\theta$ is a
substitution (i.e. a mapping $V \to (V \cup \Sigma)^*$)
then $X\theta$ is called an
instance of $X$. If $X\theta \in \Sigma^*$ we call $X\theta$ a {\em ground
instance} of $X$.\\[1ex]
We may define $P = ({\cal Y},{\cal X},\Sigma,\Sigma_0)$ where
${\cal Y}$ is a finite sets of meta-strings (the axioms) and
${\cal X}$ is a finite set of
{\em rules}, i.e. expressions of the form:
\[
{X_1 \ldots X_n\over X}
\]
where $X_1,\ldots,X_n,X$ are meta-strings such that the set of variables in $X$ is
contained in the set of variables in $X_1,\ldots,X_n$.

$\Sigma_0$ is a (non-empty) subset of $\Sigma$ (defining the strings of the
theory to be generated).
\\[1ex]
A {\em derivation $\varphi$ in $P$} is a tree such that all nodes are labelled
by strings in $\Sigma^*$. In particular:
\begin{itemize}
\item the leaves of $\varphi$ are labelled by ground instances of axioms.
\item Let $N$ be a node in $\varphi$ which is not a leaf and
    $(N,N_1),\ldots,(N,N_k)$ be the nodes from $N$ then
\[
{N_1 \ldots N_k\over N}
\]
is a ground instance of a rule in ${\cal X}$.
\end{itemize}

A proof of an $x$ in $\Sigma_0^*$ in $P$ is a derivation in $P$
with the root node labeled
by $x$. $x$ is called {\em provable} in $P$ if there exists a proof of
$x$ in $P$.\\[1ex]
{\bf fact:} as $\Sigma$ is finite there are only finitely many derivations
of length $\leq k$ for any natural number $k$, where {\em length} is the
number of symbol occurrences. Let $P[k]$ be the set of all derivations of
length $\leq k$.\\[1ex]
We prove now:\\
There is no recursive function $g$ such that for all $x \in X$:
\begin{itemize}
\item[$(*)$] $x$ is provable in $P$ iff there exists a proof $\varphi$ of $x$
    with $|\varphi| \leq g(|x|)$.
\end{itemize}
{\bf Proof:}\\
Assume that there exists a recursive $g$ such that $(*)$ holds. We construct a
decision procedure of $X$:
\[
\begin{array}{l}
\mbox{input:}\ x \in X.\\
\bullet\ \mbox{compute}\ g(|x|).\\
\bullet\ \mbox{construct}\ P[g(|x|)].\\
\bullet\ \mbox{if } P[g(|x|)] \mbox{ contains a proof of }x \mbox{ then }
    x \in X\\
\hspace*{1cm} \mbox{ else } x \not \in X.
\end{array}
\]
But we assumed $X$ to be undecidable, thus we arrive at a complete contradiction. {Q.E.D.}

It follows as a corollary that there exists no proof system which generates an undecidable
problem $X$ and $X$ is polynomially verifiable.

The result above illustrates one of the problems in acknowledging hypercomputation. Even if we have a strong hypercomputer solving, let us say, the halting problem, the verification of its correctness is ultimately unfeasible. Due to the absence of recursive bounds we cannot expect to obtain a full proof of the corresponding property (halting or non-halting) from the hypercomputer itself.

When we consider the halting problem and the property of non-halting, this can only be verified by a proof (and not by simulating a Turing machine). By the undecidability of the problem there is no complete (recursive) proof system doing the job. So when we obtain a verification from the hypercomputer concerning non-halting, the form of this verification lies outside computational proof systems.

However we might think about the following test procedure for hypercomputers:
humans create a test set of problems for an undecidable problem $X$,
i.e. a finite set $Y$ with $Y \cap X \neq \emptyset$ and $Y \cap X^c \neq \emptyset$.
The humans are in possession of the solutions,
preferably of proofs $\varphi_y$ of $y \in X$ or of $y \not \in X$ for any $y \in Y$.
This finite set may at least serve the purpose of {\em falsifying} hypercomputation
(provided the hypercomputer is not stochastic and wrong answers are admitted).
Beyond the possibility of falsification we might consider the following scenario:
the hypercomputer answers all questions concerning the test set $Y$ correctly,
and its computing time is independent of the complexity of the proofs $\varphi_y$.
Such a phenomenon would, of course, not yield a verification of the hypercomputer
but at least indicate a behavior structurally differing from computable proof systems.

But the ultimate barrier of verifying a hypercomputer is that of verifying a black box, characterized by the attempt to induce a property of infinitely many input-output pairs by a finite test set.

\section{Discussion and summary}

The considerations presented here may be viewed as special cases of a very general black box identification problem:
is it possible to deduce certain features of a black box, without screwing the box open and without knowing the intrinsic
working of the black box, from its input/output behavior alone?
Several issues of this general problem have already been discussed.
For instance, in an effort to formalize the uncertainty principle,
Moore \cite{e-f-moore} considered initial state identification problems of (given) deterministic finite automata.
Gold  \cite{go-67,blum75blum,angluin:83,ad-91,li:92} considered a question related to induction:
if one restricts black boxes to computable functions, then the rule inference problem,
i.e., the problem to find out which function
is implemented by the black box, is in general unsolvable.
The halting problem \cite{turing-36,rogers1,odi:89}
can be translated into a black box problem: given a black box
with known partial recursive  function,
then its future behavior is generally unpredictable.
Even the problem to determine whether or not a black box system is polynomial in computation space and
time appears to be far from being trivial.
In general, finding a way to falsify hypercomputation by a single counterexample
may turn out to be easier than proving hypercomputation in full generality.

So, if presented with a hypercomputer or oracle, we could only assert heuristic information, nothing more.
We have to accept the fact that more general assertions, or even proofs for computational capacities
beyond very limited finite computational capacities remain impossible, and will possibly remain so forever.

The situation is not dissimilar from claims of absolute indeterminism and randomness on a microphysical scale
\cite{svozil-qct}, where a few, albeit subtle tests of time series \cite{calude-dinneen05} generated by
quantum randomness oracles such as {\it Quantis}\textsuperscript{\texttrademark}  \cite{Quantis} can be compared against
advanced algorithmic random number generators
such as the  Rule30CA Wolfram rule 30 generator  implemented by {\it Mathematica}\textsuperscript{\textregistered}.
Beyond heuristic testing, any general statement about quantum randomness remains conjectural.

\section{Acknowledgments}
This manuscript grew out of discussions between computer scientists and physicists at the Vienna University of Technology,
including, among others, Erman Acar, Bernhard Gramlich, Markus Moschner, and Gernot Salzer.


\end{document}